\documentclass[sigconf,nonacm,screen]{acmart}
\AtBeginDocument{%
  \providecommand\BibTeX{{%
    \normalfont B\kern-0.5em{\scshape i\kern-0.25em b}\kern-0.8em\TeX}}}

\setcopyright{acmcopyright}
\copyrightyear{2018}
\acmYear{2018}
\acmDOI{XXXXXXX.XXXXXXX}

\usepackage{threeparttable}
\usepackage{float}
\usepackage{tabularx}

\acmConference[Conference acronym 'XX]{Make sure to enter the correct
  conference title from your rights confirmation emai}{June 03--05,
  2018}{Woodstock, NY}
\acmPrice{15.00}
\acmISBN{978-1-4503-XXXX-X/18/06}

\begin{document}

\title{On the Mechanics of NFT Valuation: AI Ethics and Social Media}
\subtitle{ChainScience/2023/16}

\author[Luyao Zhang*]{Luyao Zhang}
\affiliation{%
  \department{Data Science Research Center and Social Science Division}
  \institution{Duke Kunshan University}
  \country{China}
}
\authornote{Corresponding authors: \newline
Xin Tong (email: xt43@duke.edu) and Luyao Zhang (email: lz183@duke.edu) are the corresponding authors and authors of equal contributions, address: Duke Kunshan University, No.8 Duke Ave. Kunshan, Jiangsu 215316, China.}
\authornote{Also with SciEcon CIC, a not-for-profit organization aiming at cultivating interdisciplinary research and integrated talents in United Kingdom.}

\author{Yutong Sun}
\affiliation{%
  \institution{Duke Kunshan University}
  \streetaddress{Duke Ave. No.8 Kunshan}
  \city{Suzhou}
  \country{China}}
\authornote{Y. Sun, Y. Quan, and J. Cao are the students and advisees of Prof. Xintong and Prof. Luyao Zhang at HCI Blockchain Lab, Duke Kunshan University, China}

\author{Yutong Quan}
\affiliation{%
  \institution{Duke Kunshan University}
  \streetaddress{}
  \city{}
  \country{China}}
\authornotemark[3]

\author{Jiaxun Cao}
\affiliation{%
  \institution{Duke Kunshan University}
  \streetaddress{}
  \city{}
  \country{China}}
\authornotemark[2]

\author{Xin Tong}
\affiliation{%
  \institution{Duke Kunshan University}
  \streetaddress{}
  \city{}
  \country{China}}
\authornotemark[1]

\renewcommand{\shortauthors}{Luyao Zhang, et al.}

\begin{abstract}
As CryptoPunks pioneers the innovation of non-fungible tokens (NFTs) in AI and art, the valuation mechanics of NFTs has become a trending topic. Earlier research identifies the impact of ethics and society on the price prediction of CryptoPunks. Since the booming year of the NFT market in 2021, the discussion of CryptoPunks has been propagated on social media. Still, existing literature hasn't considered the social sentiment factors after the historical turning point on NFT valuation. In this paper, we study how sentiments in social media, together with gender and skin tone, contribute to NFT valuations by an empirical analysis of social media, blockchain, and crypto exchange data. We evidence social sentiments as a significant contributor to the price prediction of CryptoPunks.
Furthermore, we document structure changes in the valuation mechanics before and after 2021. Although people's attitudes towards Cryptopunks are primarily positive, our findings reflect imbalances in transaction activities and pricing based on gender and skin tone. Our result is consistent and robust, controlling for the rarity of an NFT based on the set of human-readable attributes, including gender and skin tone. Our research contributes to the interdisciplinary study at the intersection of AI, Ethics, and Society, focusing on the ecosystem of decentralized AI or blockchain. We provide our data and code for replicability as open access on Harvard Dataverse and GitHub. 
\end{abstract}





 % \includegraphics[width=\textwidth]{sampleteaser}
  %\caption{Seattle Mariners at Spring Training, 2010.}
  %\Description{Enjoying the baseball game from the third-base
  %seats. Ichiro Suzuki preparing to bat.}
  %\label{fig:teaser}

%\received{}
%\received[]{}
%\received[]{}

\maketitle

\section{Introduction}
\label{sec:intro}

Recent years have seen the rise of NFTs that refer to digital records stored on blockchains, e.g., Ethereum, Solana, Tezos, etc. With blockchain technology, NFTs can be traded with cryptographic mechanisms to reach a consensus on the recorded transaction data among untrusted users ~\cite{sharma_its_2022, garay2017bitcoin,nakamoto2008bitcoin}. In addition, NFTs ensure the authenticity and tamper-resistance of the transacted assets ~\cite{sharma_its_2022,dyson2019challenges}. When trading NFTs, the transactions are programmed into a smart contract that stores the link to the traded assets in the metadata so that the assets can be identified only by the unique token ID and address of the parent contract ~\cite{kapoor_tweetboost_2022}. Furthermore, blockchain technology protects the anonymity of NFT owners by excluding interventions, regulations, and censorship from centralized institutions such as governments and banks ~\cite{sharma_its_2022}. These benefits of NFTs were largely recognized by the public starting in 2021 as NFT artworks went viral ~\cite{kapoor_tweetboost_2022}. In addition to digital art, NFTs also represent ownership of music, game objects, collectibles ~\cite{nguyen_racial_2022, sharma_its_2022}, etc. Currently, the largest NFT collection on the market is CryptoPunks which contains a set of pixel art images ~\cite{noauthor_CryptoPunks_nodate}. The rise of NFT digital art has led to the emergence of NFT artworks in traditional auction houses ~\cite{kapoor_tweetboost_2022}. For instance, in 2021, a digital artist named Beeple sold his work ‘Everydays: The First 5000 Days’ for over \$69 million.

As NFT communities continue to grow, there has been an emergent academic agenda in analyzing NFT marketplaces. ~\cite{kapoor_tweetboost_2022}, especially on NFT valuation and factors that influence NFT prices ~\cite{dowling2022non, ante2022non, dowling_fertile_2022, chohan_nft_2021}. For instance, researchers have identified sales histories and visual features, such as gender and skin tone, as predictive features of NFT prices ~\cite{nguyen_racial_2022, nadini_mapping_2021}. Such discoveries have led to discussions on the ethics in the NFT market, e.g., potential racial biases such as CryptoPunks with lighter skin tones being sold at higher prices, even after controlling for the rarity and market conditions ~\cite{nguyen_racial_2022}. ~\footnote{Besides racial and gender bias, the aforementioned anonymity and decentralization of NFT transactions have also raised concerns about illegal money laundering ~\cite{Faraj_governmentonnfts_2023}.} In the meantime, social media plays a critical role in NFT transaction activities. Social media platforms such as Twitter and Reddit have been the main channels for the public to learn about NFT-related content ~\cite{luo_understanding_2022}, including relevant events for NFT projects ~\cite{luo_understanding_2022}, NFT artwork promotion by creators ~\cite{chohan_nft_2021} and so forth.\footnote{Prior research also reveals that social media has driven the formation and development of NFT communities ~\cite{luo_understanding_2022}. In addition, NFT-related activities on social media have been suggested to be an important feature in predicting NFT prices ~\cite{kapoor_tweetboost_2022, luo_understanding_2022, jain2022nft}. For instance, prior work that compared the predictive power of image features and social media features of NFTs suggests that social media features are more influential to NFT prices than the features of NFT products 
~\cite{kapoor_tweetboost_2022}. Among the social media features, public opinions have been suggested to be strong indicators of valuation. Besides being a strong predictive indicator, social media activities also affect purchasing behavior and influence NFT prices ~\cite{efendioglu_rise_2022}. For example, both the original NFT creator and buyers can promote their assets on social media, attracting other potential buyers ~\cite{chohan_nft_2021}.} Since the boom of the NFT market in 2021, the discussion of CryptoPunks has propagated on social media. Nevertheless, the literature has not considered the social sentiment factors after historical turning points in NFT valuation. Moreover, although 2021 has witnessed explosive growth in NFT markets, existing analysis rarely includes data beyond 2021. Taking the largest NFT collection, CryptoPunks, as an example, we seek to investigate the following research questions (RQs) in this study:

\begin{itemize}
   \item {\textbf{RQ1:}} What are the characteristics of tweet sentiment, skin tone, and gender of CryptoPunks? (\ref{sec:rq1})
   \item {\textbf{RQ2:}} How can sentiment in social media, together with gender and skin tone, contribute to NFT valuation?  (\ref{sec:rq2})
   \item {\textbf{RQ3:}} What are the structural changes in CryptoPunks valuation after the explosive growth of NFT markets in 2021? (\ref{sec:rq3})
 \end{itemize}

 To answer RQ1, we first conducted sentiment analysis on CryptoPunk-related tweets using the Valence Aware Dictionary for Sentiment Reasoning (VADER). The results suggest that the discussions on CryptoPunks on Twitter from 2017 to 2022 are overall positive. Since the explosive growth in late 2021, the discussions have become even more positive and stable. We also examined the skin tone and gender characteristics of CryptoPunk-related transactions and tweets, revealing the imbalances of transactions and tweets about different genders and skin tones of CryptoPunks. To answer RQ2, we evaluated the predictability of tweet sentiments by comparing a prediction model including sentiment scores as an independent variable and one without controlling gender, skin tones, and other technical indicators. The results suggest that tweet sentiment significantly impacts the prediction of CryptoPunk prices without reducing the predictability of other predictors (e.g., gender and skin tone). To answer RQ3, we compared the empirical data analysis from 2017 to 2021 and 2021 to 2022. The results indicate a structural change in CryptoPunk valuation mechanics after 2021. For gender, we find a reversal of the previous trend, with males now predicted to have higher prices than females. For skin tones, the statistical significance of light skin is wholly lost, and the statistical significance of medium skin on price prediction is significantly reduced. The nonhuman skin tone and sentiment scores still contribute positively to pricing.  Our result is consistent and robust, controlling for the rarity of an NFT based on the set of human-readable attributes, including gender and skin tone.

\section{Background and Related Work}
\label{sec:related}

\subsection{NFTs, Marketplaces, and Ethics}
\label{blockchain}

\subsubsection{The Background of NFTs and Marketplaces}
NFTs refer to digital records stored on blockchains (e.g., Ethereum, Solana and Tezos ~\cite{nguyen_racial_2022}), which are a technology that ensures a consensus on the recorded transaction data among untrusted users with cryptographic mechanisms ~\cite{sharma_its_2022, garay2017bitcoin, nakamoto2008bitcoin}. NFTs represent ownership of pieces of content, i.e., assets, such as digital art, music, game objects, and collectibles ~\cite{nguyen_racial_2022, sharma_its_2022}. These assets are traded with cryptocurrencies through online marketplaces, e.g., OpenSea, Rarible, LooksRare, etc ~\cite{kapoor_tweetboost_2022, nguyen_racial_2022, sharma_its_2022}. When trading NFTs, the transactions and ownership of the asset are written in a smart contract, where the link to the asset is stored in its metadata ~\cite{kapoor_tweetboost_2022}. Through this process, each asset can be identified by the parent contract's unique token ID and address, ensuring its authenticity and tamper-resistance ~\cite{sharma_its_2022, dyson2019challenges}. Additionally, since blockchain technology can exclude interventions, regulations, and censorship from centralized authorities such as banks and governments during transactions ~\cite{sharma_its_2022}, blockchain technology can protect the anonymity of NFT owners. 

As the public recognizes the benefits of NFTs, recent years have seen explosive growth in the NFT market. Currently, the largest NFT collection (i.e., a set of NFT assets on the marketplaces with similar traits and properties ~\cite{kapoor_tweetboost_2022}) is CryptoPunks, a series of pixel art images that contain punky-looking men, women, and nonhuman creatures such as apes and aliens ~\cite{noauthor_CryptoPunks_nodate}. Created in 2017 by Larva Labs ~\cite{nguyen_racial_2022}, the CryptoPunks collection surpassed \$1 billion in sales in 2021 ~\cite{pinto-gutierrez_nft_2022}. Not only CryptoPunks but also the entire NFTs have received phenomenal attention in 2021, leading to the proliferation of NFT digital art in traditional auction houses ~\cite{kapoor_tweetboost_2022}. For example, ‘Everydays: The First 5000 Days’, an NFT artwork by Beeple, was sold for over \$69 million on March 11, 2021. In November 2021, the average sales price of NFTs reached its highest value of USD 953.44 ~\cite{noauthor_nfthist_nodate}. Nevertheless, according to a market report ~\cite{CHAINALYSIS_transactionsince2021_2022}, after reaching a peak in late 2021, the growth of NFT transaction activities has stabilized. While this report has not analyzed the reasons behind such changes in NFT transaction activities, a previous study that analyzed the rise and fall of a widely recognized blockchain game---CryptoKitties---suggests that the fall of the game was due to (1) the oversupply of game assets---kitties, (2) the decrease in player income, (3) a widening gap between rich and poor players, and (4) the limitations of blockchain systems ~\cite{jiang2021cryptokitties}.

The rise of NFTs and NFT artworks has also led to academic discussions, mainly on NFT market analysis. A body of prior work has focused on analyzing the relationship between blockchain, cryptocurrencies, and NFTs ~\cite{kapoor_tweetboost_2022}, e.g., understanding the correlation between the entire NFT market and submarkets (e.g., Ethereum and Bitcoin) that can have millions of dollars traded every day ~\cite{ante2022non, dowling2022non}. For instance, Karim et al. ~\cite{karim2022examining} examined the interrelatedness of NFTs, DeFi tokens, and cryptocurrencies. Their results suggest that blockchain markets have strong volatility connectedness at the median, extremely low, and extremely high volatility conditions, revealing a higher diversification of NFTs compared to DeFis and Cryptos. 

In addition to general NFT market analysis, the literature has discussed NFT asset valuation ~\cite{kapoor_tweetboost_2022, nadini_mapping_2021}, factors determining the prices of NFTs ~\cite{dowling2022non, ante2022non, dowling_fertile_2022, chohan_nft_2021}, etc. For example, Horky et al. ~\cite{horky2022price} classified factors that influence NFT asset pricing into two types: (1) endogenous factors, including some common terms in the description of NFTs, and (2) exogenous factors, including the size of NFTs in bytes, and the format of NFTs (e.g., jpeg, png, gif, and mp4). Similarly, Maouchi et al. ~\cite{maouchi2022understanding} propose a set of endogenous predictors of NFT prices (e.g., the traded volume and the total value locked expressed in ETH), and exogenous predictors (e.g., investors' sentiment reflected from Google Trends searches for different crypto assets tickers and financial factors such as the economic policy uncertainty index).

\subsubsection{Ethical Concerns of NFTs}
Notably, both endogenous and exogenous factors have raised ethical concerns. For instance, endogenous factors such as the anonymity and decentralization of NFTs have been used to launder money by converting it into a digital asset ~\cite{Faraj_governmentonnfts_2023}. To address this concern, especially since 2022, governments have been taking action to regulate NFT transactions to ensure the same anti-money laundering regulations as traditional financial transactions ~\cite{Faraj_governmentonnfts_2023}.

Some endogenous predictors, such as visual features of NFTs, suggest racial and gender bias in NFTs. As Nadini et al. ~\cite{nadini_mapping_2021} suggest, sale history and visual features are two predictors of NFT prices with reasonable predictability. Gender (e.g., female and male), skin tone (e.g., dark, medium, and light), and species (e.g., human and nonhuman) are prominent visual features of NFTs. Therefore they can influence NFT valuation. For example, one previous study that focused on the skin tones of NFTs suggests that CryptoPunks with lighter skin tones are sold at higher prices, even after controlling for the rarity and market conditions, meaning potential racial bias in the NFT market ~\cite{nguyen_racial_2022}.

While studies on discrimination in NFT transactions are largely scant, prior work on racial discrimination in physical collectibles suggests multiple sources of racial discrimination in prices, including suppliers, customers, and other parties ~\cite{becker2010economics}. For customer racial discrimination, studies that have examined gender and racial influences on prices have mixed and contradictory findings ~\cite{nguyen_racial_2022, mcgarrity1999consumer, primm2011investigating, nardinelli1990customer}. For instance, Nardinelli and Simon \cite{nardinelli1990customer} suggest that the player's race on the baseball cards can affect the consumer's desired price. On the other hand, other studies have reported no or insignificant evidence of racial discrimination among consumers in the case of the baseball card market ~\cite{mcgarrity1999consumer, primm2011investigating}. However, whether ethics-related endogenous predictors such as gender and skin tones affect NFT valuation remains unknown.

In summary, while these strands of work have analyzed the NFT market from the employed technologies and NFT valuation with ethical factors, they ignore the perspective of social media that can help predict and influence the prices of NFT assets ~\cite{kapoor_tweetboost_2022}, unveiling market behaviors through mineable information from social media communities ~\cite{luo_understanding_2022}. Hence social media analysis can help better understand the NFT market and valuation.

\subsection{Social Media Analysis on NFT-related Content}
\label{sec:socialmedia}

Social media has contributed mainly to the rise of NFTs, as mainstream social media platforms (e.g., Twitter, Reddit, and Discord) have been the main channels for the public to learn about some NFT-related content such as new events for NFT projects ~\cite{luo_understanding_2022}. Previous studies have found that social media has driven the development of NFT communities ~\cite{luo_understanding_2022}. For instance, Casale-Brunet et al. ~\cite{casale-brunet_impact_2022} analyze the NFT communities on Twitter using social network analysis (SNA), suggesting that most top NFTs could be considered single communities, in which most projects are influenced by the development of the Bored Ape Yacht Club ~\cite{opensea_1_nodate} collection.

Furthermore, a large body of work has indicated that social media activities are essential in predicting NFT prices ~\cite{kapoor_tweetboost_2022, luo_understanding_2022, jain2022nft}. For example, Kapoor et al. ~\cite{kapoor_tweetboost_2022} find that social media features improve an NFT valuation classification task. Additionally, two preliminary studies were conducted to analyze the relationship between Twitter trends and the average price of tokens in an NFT collection ~\cite{casale-brunet_impact_2022, jain2022nft, kapoor_tweetboost_2022}. Before NFT valuation analysis, a study explored the interaction between social media and cryptocurrencies, suggesting that the social media indicator drives predictions for cryptocurrency prices ~\cite{ortu2022technical}. Researchers who have compared the predictive power of image features of NFTs and social media features have found that social media features have a more substantial impact on NFT prices than the NFT product itself ~\cite{kapoor_tweetboost_2022}.

Fu et al.\cite{fu2022ai} show that tweets discussed profound topics of ethical concern in the blockchain ecosystem, including security, equity, and emotional sentiments. Among different social media predictors, public sentiment has been regarded as a prominent predictive feature for prices and returns ~\cite{sul2017trading, broadstock2019social, nguyen2015sentiment, lamon2017cryptocurrency, wolk2020advanced}. In the stock market, researchers found that tweet sentiment toward a firm's stock and the broader financial market plays a vital role in stock returns ~\cite{sul2017trading, broadstock2019social}. For example, Sul et al. ~\cite{sul2017trading} have analyzed the correlation between the sentiment of 2.5 million tweets of approximately 500 firms and the stock returns of these firms. Their results suggest that tweet sentiments from users with fewer than 171 followers that were not retweeted had the most significant impact on future stock returns. Another study built a model to predict stock prices using social media sentiment ~\cite{nguyen2015sentiment}. Unlike research that considered the overall sentiment, their approach incorporated the sentiments of specific topics to build the stock price prediction model, which showed a better performance. More recently, researchers have started examining sentiment analysis to predict prices of bitcoin and other cryptocurrencies ~\cite{wolk2020advanced, lamon2017cryptocurrency}. Compared to stock prices, the decentralization of cryptocurrencies has made people's attitudes more critical than institutional regulations regarding price predictions ~\cite{wolk2020advanced}. As a result, previous studies generally suggest that people's sentiments reflected in social media and web searches have significantly influenced cryptocurrency prices ~\cite{wolk2020advanced, lamon2017cryptocurrency}. However, studies on sentiment analysis as a computational tool to predict NFT prices are still largely scarce.

In addition to being a predictive feature, social media interactions have been suggested to significantly affect purchasing intention ~\cite{efendioglu_rise_2022}, hence influencing NFT prices. For example, the original NFT creator can promote their works on social media and buyers can resell NFTs to other consumers on social media ~\cite{chohan_nft_2021}. Twitter recently introduced official support for using NFTs as profile pictures ~\cite{meyns_what_2022}, attracting more people to join NFT-related conversations. However, the topics covered in these conversations and the attitudes toward NFT on social media have been largely unexplored. Examing NFT-related content via social media analysis can provide scope for predicting NFT prices and better understanding market behaviors. To this end, we conducted a study to explore the topics related to CryptoPunks, the largest NFT collection to date, on Twitter which accounts for more than 70\% of the total traffic from social media on OpenSea \footnote{https://opensea.io/assets/ethereum/ 0xbc4ca0eda7647a8ab7c2061c2e118a18a936f13d/1}.

\section{Data and Methods}
\label{sec:method}

\subsection*{Data and Code Availability Statement}
\label{sec:datasource}
The data and code for replicability of this study are released as open access on Harvard Dataverse\cite{DVN/YMZC30_2023} and GitHub\footnote{https://github.com/HCI-Blockchain/NFT-2023}.

\subsection{Data Collection}
\label{sec:datacollection}

To conduct a comprehensive sentiment analysis of CryptoPunks and investigate the influence of sentiment and ethical factors on their price, we collected social media data and CryptoPunk transaction data. Our study timeframe ranges from June 23, 2017, the release date of CryptoPunks, to October 31, 2022, covering a long period and including 2021, a year of significant change for the NFT market. This timeframe accurately reflects the dynamic development and changes in social media discussions about CryptoPunks, enabling us to investigate possible structural changes.

\subsubsection{Social Media Data}
\label{sec:mediadata}

We used Snscrape API\footnote{https://github. com/JustAnotherArchivist/snscrape} to gather Twitter data related to CryptoPunks, which we divided into two datasets. The first set includes all CryptoPunks-related tweets using "cryptopunk" as the keyword, and we screened for English-language tweets to obtain 122,379 tweets. The second set of Twitter data consists of tweets related to gender and skin tone in CryptoPunks. We used keywords such as "female", "male", "dark", "light", "medium", "albino", "alien", "ape", and "zombie" as filters to obtain a total of 5,883 tweets discussing gender or skin tone.

Using the first set of Twitter data, we calculate the daily tweet volume to capture the overall change in the volume of social media discussion about CryptoPunks from 2017 to 2022. We then perform sentiment analysis to compute daily tweet sentiment scores to comprehend the changes in users' emotional attitudes toward CryptoPunks, as discussed in detail in \ref{sec:sentiment}. The CryptoPunk Twitter sentiment score is also utilized as a predictor in subsequent price prediction studies in \ref{sec:hedonic}.

With the second data set, we investigated potential gender and racial discrimination issues in social media discussions about CryptoPunks. We measured the discussion volume of each ethical topic by calculating the word frequency of gender- and skin-tone-related keywords. Additionally, we calculated separate sentiment scores for tweets related to each ethical keyword to compare the differences in user attitudes toward different genders and skin tones. These findings will be discussed in \ref{sec:rq1}.

\subsubsection{CryptoPunks Transaction Data}
\label{sec:transactiondata}
The study utilizes transaction data sourced from Dune Analytics \footnote{https://dune.com/browse/dashboards}, a powerful tool designed for querying transaction data on blockchain and Ethereum. The Dune API is leveraged to retrieve transaction data for CryptoPunks from 2017 to 2022. The original dataset consists of 13 variables: skin tone, gender, price (in ETH), and trading wallet address. These variables are crucial in computing daily sales volume change and daily active wallet change data. Additionally, data on Ethereum gas prices, which reflect the transaction costs, are sourced from Etherscan\footnote{https://etherscan.io/}, a block explorer and analytics platform for Ethereum. The ETH/USD exchange rate is from investing.com\footnote{https://www.investing.com/}. The datasets are merged by date and cleaned to ensure they are ready for processing. Combining the transaction data with Twitter sentiment data can contribute to the hedonic regression model for predicting CryptoPunk prices.

\subsection{Sentiment Analysis}
\label{sec:sentiment}

Sentiment analysis is a natural language processing technique used to identify, extract, and quantify subjective information from text ~\cite{pang2008opinion}. It involves analyzing the underlying sentiment or emotional tone of a piece of text, such as a tweet or a news article, and categorizing it as positive, negative, or neutral ~\cite{liu2012sentiment}. Sentiment analysis can be applied to various types of text data, including social media posts, customer reviews, and news articles, to gain insights into people's opinions and attitudes toward a particular topic or entity ~\cite{hutto_vader_2014}. 

Our sentiment analysis is performed using the VADER (Valence Aware Dictionary and Sentiment Reasoner) sentiment analysis tool, which is a rule-based model for sentiment analysis of social media text ~\cite{hutto_vader_2014}. VADER uses a lexicon that maps lexical features to sentiment scores, which are then combined to produce a compound sentiment score for a given piece of text ~\cite{hutto_vader_2014}. The compound score ranges from -1 to 1, with -1 indicating extremely negative sentiment and 1 indicating extremely positive sentiment, while scores close to zero indicate neutral sentiment ~\cite{hutto_vader_2014}. We apply VADER to each CryptoPunk-related tweet in our dataset to obtain a compound sentiment score for that tweet.

To calculate the daily sentiment score, we aggregate the sentiment scores of all CryptoPunk-related tweets posted on that day, and then divide the sum by the total number of tweets posted on that day. This gives us an average daily sentiment score that represents the overall emotional attitude toward CryptoPunks on that day. We repeat this process for each day in our first social media dataset to obtain a time series of daily sentiment scores for CryptoPunks-related tweets. These daily sentiment scores are then used in the subsequent analysis to explore the relationship between sentiment and price.

In addition, we also calculate sentiment scores for tweets related to each gender and skin tone keyword. To achieve this, we first filter tweets containing each keyword and then calculate the sentiment score for each tweet in the second Twitter dataset using the VADER sentiment analysis tool. Finally, we compute the average sentiment score for all tweets related to each keyword. These sentiment scores for each keyword provide insights into the emotional attitudes toward different genders and skin tones in social media discussions about CryptoPunks.

\subsection{Hedonic Regression}
\label{sec:hedonic}

Our study aims to compare the effects of gender and skin tone on the prediction of CryptoPunks' price using the Hedonic regression model. This model estimates the value of a product by evaluating the value of its composing parts \cite{rosen_1974}. A previous study by Nguyen \cite{nguyen_2022} analyzed the impact of gender and skin tone on CryptoPunks' price using the Hedonic regression model and market transaction data, with rarity, active market wallets, CryptoPunk sales, and the ETH/USD exchange rate as control variables, where the rarity of an NFT is based on the set of human-readable attributes, including gender and skin tone. In our study, we build on Nguyen's model by adding a new independent variable---the sentiment score of CryptoPunk-related tweet data, which reflects people's attitudes toward the market \cite{cheng_wicks_2014, kapoor_tweetboost_2022}. We compare the regression results of our updated hedonic regression model with Nguyen's results to see if the sentiment score has any effect on CryptoPunks' pricing prediction.

In addition, we expand the time span of the data used in Nguyen's study from 2017-2021 to 2017-2022, to explore whether the explosive growth of NFT markets in late 2021 had any impact on CryptoPunks' pricing mechanics. To conduct a more comprehensive analysis, we run regressions for three time periods: 2017-2021, 2021-2022, and 2017-2022. By comparing the results of our updated model with Nguyen's analysis, we investigate whether any structural changes occurred in the valuation mechanism of the CryptoPunks market before and after 2021.

In the hedonic regression model, we use the CryptoPunks transaction data to estimate the coefficients of the model \cite{nguyen_racial_2022}. These coefficients allow us to examine the influence of each gender and skin tone type on CryptoPunks' price prediction. Since gender and skin tone are categorical variables, we set them as dummy X variables with values of 0 or 1. The hedonic regression formula is represented as follows:
 
\begin{equation}
\label{hedonic1}
y=b_0+b_1X_{Dark}+b_2X_{Light}+b_3X_{Male}+b_4X_{Medium}
\end{equation}

\begin{equation}
\label{hedonic2}
+b_5X_{Non_human}+b_6X_{Rarity}+b_7X_{ETH/USD}
\end{equation}

\begin{equation}
\label{hedonic3}
+b_8X_{Sales}+b_9X_{Gas price}+b_{10}X_{Sentiment}+b_{11}X_{Active wallet}
\end{equation}

We use a hedonic regression model with skin tone, gender of CryptoPunks, and sentiment score as independent variables $X$, and log price (in USD) as the dependent variable $y$. To meet the prerequisite for using time series data in the regression model, we transform the price into the log price as it passes the stationary test, unlike the price. We set Female and Albino as the basic variables, which means that all the coefficients $b$ before gender and skin tone $X$ are relative to the Female and Albino skin tone type. For instance, $b_{0}$ represents the expected price when the CryptoPunks have Male and Albino skin, and $b_{1}$ is the number of increases (or decreases) compared to $b_{0}$ in Log price when the CryptoPunks are Female and have Dark skin. A positive $b_{1}$ value indicates that Female CryptoPunks with Dark skin are expected to have a higher value than Female CryptoPunks with Albino skin. Similarly, $b_{3}$ represents the number of increases (or decreases) compared to $b_{0}$ in log price prediction when the CryptoPunks are Male and have Dark skin. A positive $b_{3}$ means that Male CryptoPunks with Dark skin are expected to have higher values than Female CryptoPunks with Dark skin. A similar explanation applies to other coefficients $b$.

\section{Results and Discussion}
\label{sec:results}

\subsection{Twitter Sentiment, Gender, and Skin Tone for CryptoPunks}
\label{sec:rq1}

To comprehensively understand social media sentiment and possible gender and skin tone discrimination situations among CryptoPunks, we conducted an integrated analysis combining gender and skin tone information from social media data and transaction data.

\subsubsection{Twitter Sentiment Analysis}
\label{sec:rqsentiment}

\begin{figure}[ht]
    \centering
    \includegraphics[width=0.4\textwidth]{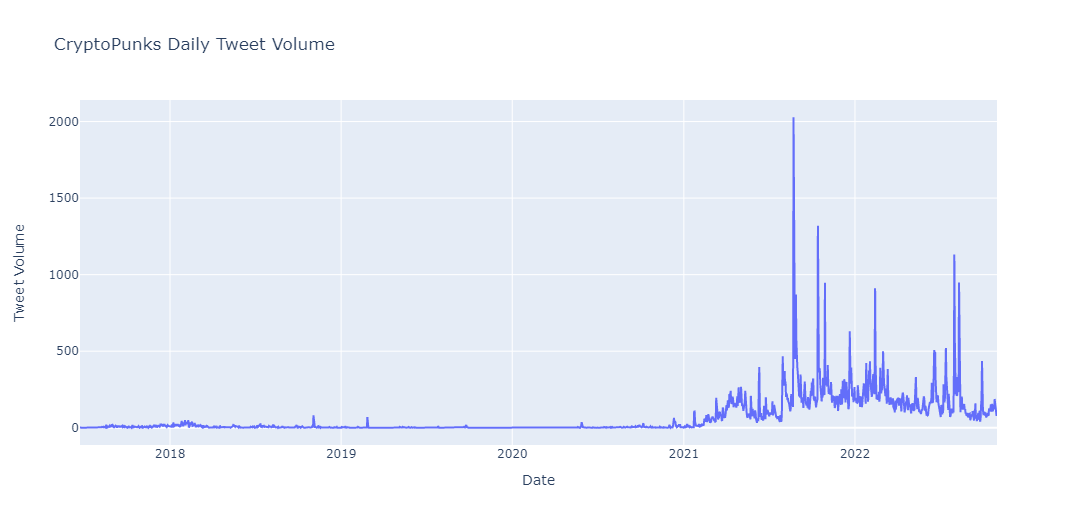}
    \caption{CryptoPunks daily tweet volume.}
    \label{fig:volume}
\end{figure}

We visualized the change in the daily tweet volume of CryptoPunks from 2017 to 2022 (see \ref{fig:volume}) and found that the discussion volume of CryptoPunks on Twitter has increased significantly since 2021, even reaching 2,028 tweets at the peak. This surge in social media discussion is related to the boom of NFTs in 2021.

\begin{figure}[t!]
    \centering
    \includegraphics[width=0.4\textwidth]{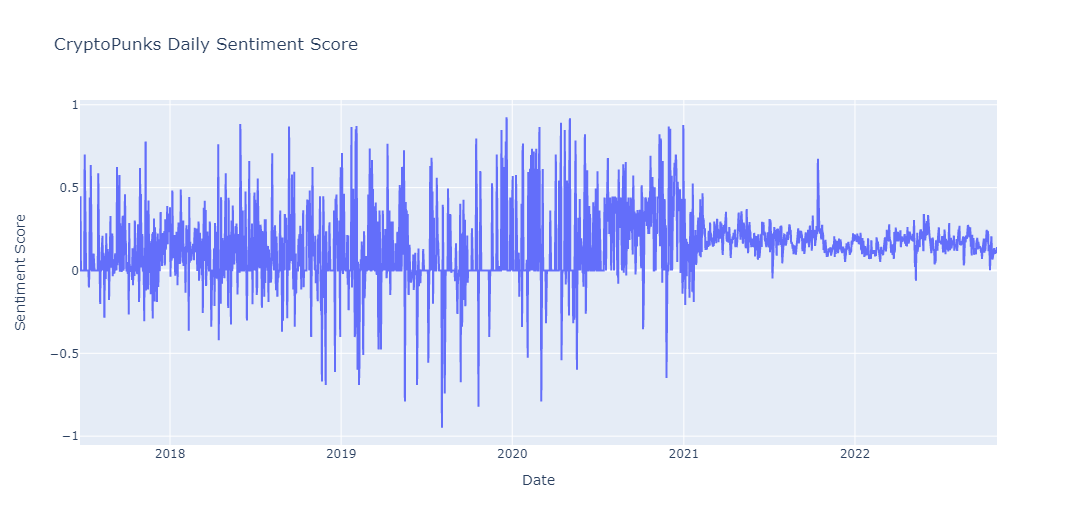}
    \caption{CryptoPunks daily sentiment score.}
    \label{fig:sentiment}

    \includegraphics[width=0.4\textwidth]{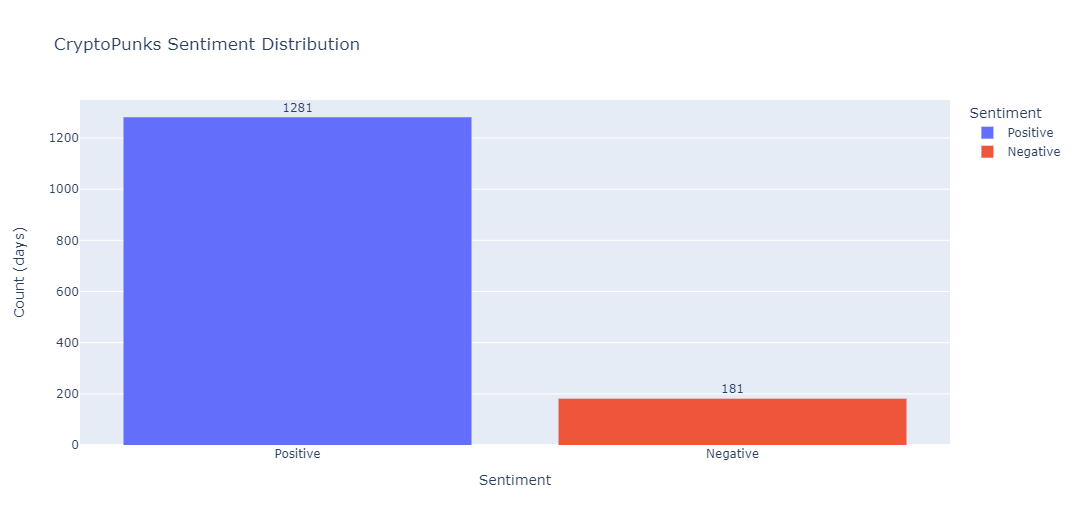}
    \caption{CryptoPunks sentiment distribution (days).}
    \label{fig:distribution}
    
\end{figure}

The sentiment analysis of all CryptoPunk-related tweets showed two main findings. First, from 2017 to 2022, the discussion of CryptoPunks by social media users was volatile but mainly positive. For example, \ref{fig:sentiment} is a plot of the change in the daily sentiment score of CryptoPunks, where the data fluctuate between -1 and 1, above 0 represents positive sentiment, and below 0 represents negative sentiment. \ref{fig:distribution} provides a clear picture of the sentiment distribution, with blue representing positive tweets and red representing negative tweets, showing far more days with positive tweets than with negative tweets. Second, after 2021, the sentiment of tweets about CryptoPunks was smoother and more dominantly positive. According to \ref{fig:sentiment}, we can see that before 2021, the sentiment of tweets fluctuated wildly between -1 and 1, but after 2021, the sentiment score was generally stable between 0 and 0.5 and showed a more upbeat performance. This change might be related to the significant increase in the inclusiveness of NFT markets after 2021.

\subsubsection{Gender and Skin Tone Analysis of Transaction Data}
\label{sec:rqgs1}

\begin{figure}[ht]
    \centering
    \includegraphics[width=0.4\textwidth]{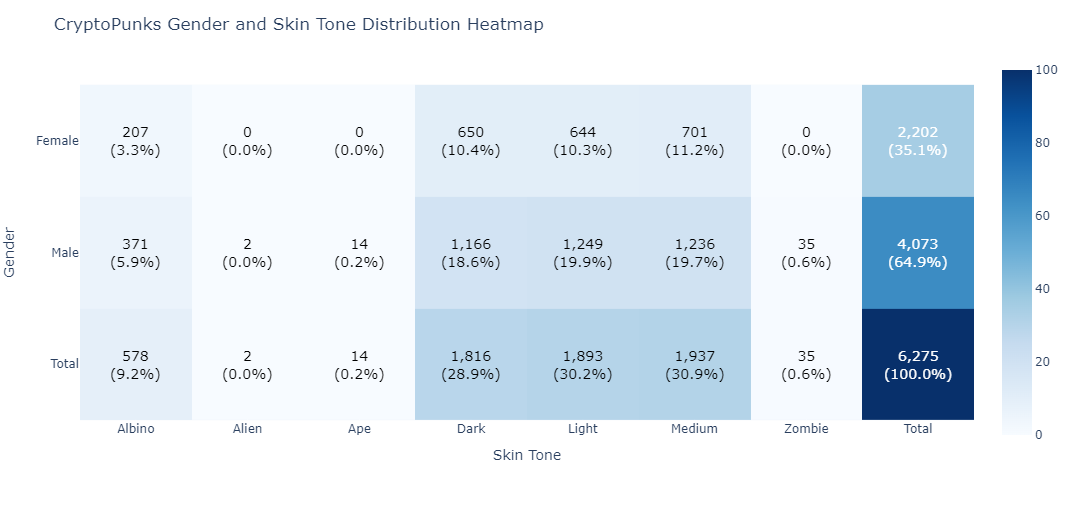}
    \caption{CryptoPunks gender and skin tone distribution heatmap.}
    \label{fig:heatmap}
\end{figure}

We calculate the number and proportion of CryptoPunks traded from 2017 to 2022 according to gender and skin tone information in the transaction data. According to the statistics, CryptoPunks have two genders, Male and Female, and five skin tones, Dark, Medium, Light, Albino, and Nonhuman. Among them, Nonhumans are divided into three types, namely Alien, Ape, and Zombie. Here, we include the three Nonhuman variables separately in the statistics. \ref{fig:heatmap} shows the gender and skin tone distribution of traded CryptoPunks. Of the 6,275 CryptoPunks traded in total, Male CryptoPunks account for 64.9\%, while Female CryptoPunks account for only 35.1\%. Moreover, for each skin tone of CryptoPunks, the number of Males is higher than that of Females. The skin tone distribution showed that Nonhuman types are far less common than Human types, among which Dark, Light, and Medium skin tones all account for approximately 30\%, while only 9.2\% are Albino skin. The rarity of an NFT based on the set of human-readable attributes, including gender and skin tone as in \ref{fig:heatmap}, is further controlled in our follow-up analysis.

\subsubsection{Twitter Discussion Analysis of Ethical Keywords}
\label{sec:rqgs2}

\begin{figure}[ht]
    \centering
    \includegraphics[width=0.4\textwidth]{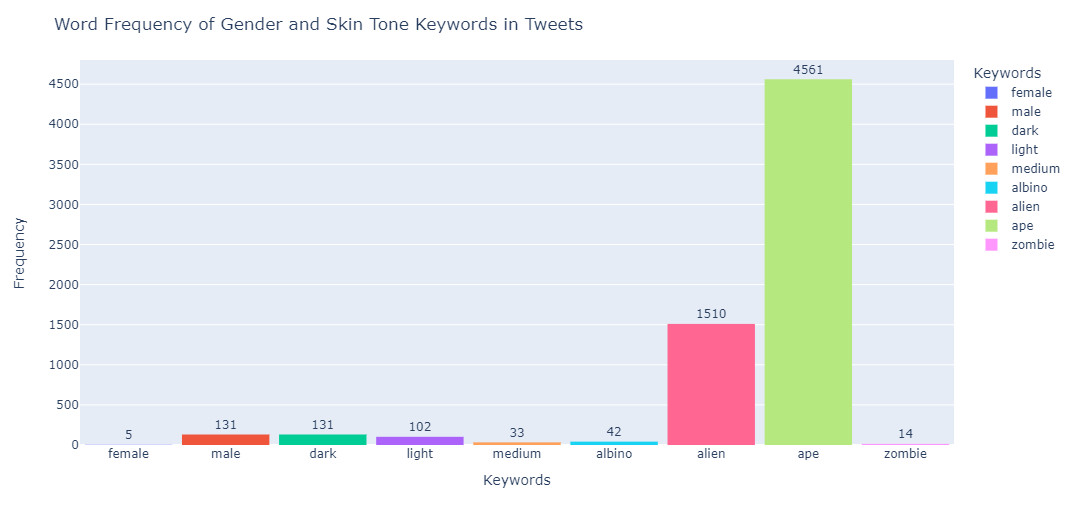}
    \caption{Word frequency of gender and skin tone keywords in tweets.}
    \label{fig:word}
\end{figure}
We conduct an analysis of ethical keywords in CryptoPunks tweets, with a focus on gender and skin tone. \ref{fig:word} presents the word frequency distribution of these keywords. Surprisingly, the majority of discussions on Twitter revolve around Nonhuman types, particularly Alien and Ape skin. This finding aligns with a prior study on CryptoPunks \cite{nguyen_racial_2022}, which attributes the trend to the rarity and greater visual appeal of Nonhuman types. Regarding the Human types, the most attention is given to Dark skin, followed by Light skin. Medium and Albino skin receive comparatively less attention. The gender gap is quite significant, with Male CryptoPunk discussion dominating at 131 instances while Female CryptoPunks appear only six times. Contrary to our hypothesis, the scarcer Female CryptoPunks harvest less yield a lower discussion volume.

\begin{figure}[ht]
    \centering
    \includegraphics[width=0.4\textwidth]{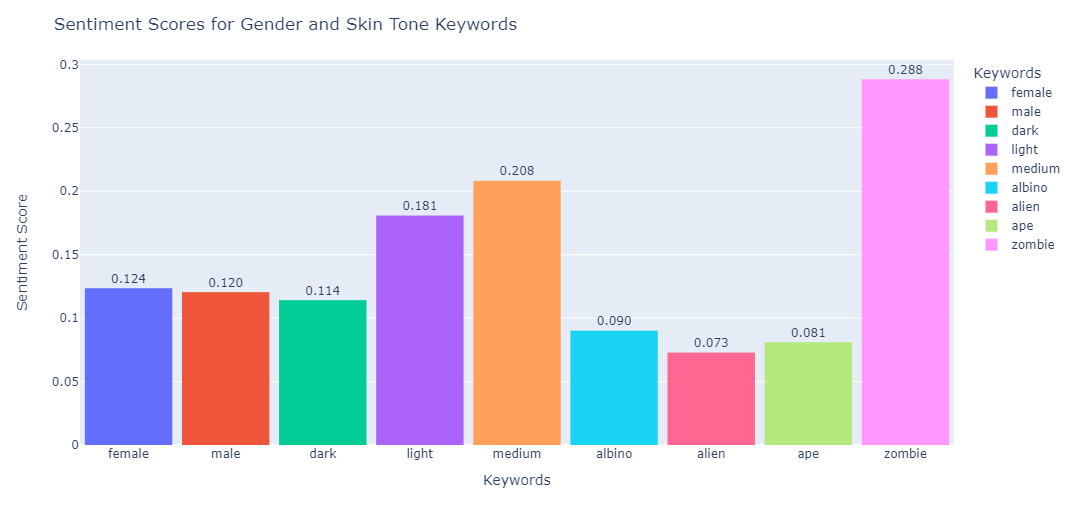}
    \caption{Sentiment Scores for Gender and Skin Tone Keywords}
    \label{fig:gssentiment}
\end{figure}

\ref{fig:gssentiment} shows the tweet sentiment for each keyword. First, corresponding to the previous results of the sentiment analysis for all tweets, the sentiment scores are positive for all keywords. Second, we find that Zombie, Medium, and Female have high sentiment scores, but considering the small number of tweets containing these keywords, sentiment scores may be biased and not necessarily significant for reference. Third, although Alien and Ape are widely discussed, people's attitudes toward them are less positive than those toward other Human species. Additionally, regarding the skin tone of the Human species, people hold more positive attitudes toward Light than Dark skin. Albino received the lowest positive score in the limited discussion. Finally, the difference in sentiment scores between Male and Female CryptoPunks was small, but it is unclear whether people have the same attitudes toward Male and Female CryptoPunks due to the limited amount of data.

From 2017 to 2022, social media discussions about CryptoPunks generally have a positive trend, which becomes more stable and positive following a surge in Twitter volume after 2021. However, an analysis of Twitter discussions and CryptoPunks trading reveals imbalances in terms of gender and skin tone. Specifically, Male CryptoPunks receive more attention than Female CryptoPunks, and typical skin tones receive more attention than Albino skin.

Previous analysis of CryptoPunk pricing indicates that Light and Albino CryptoPunks are traded at significantly higher prices, while Dark CryptoPunks are traded at lower prices \cite{nguyen_racial_2022}. Our sentiment analysis of CryptoPunks' skin tones on Twitter confirms that users generally have a more positive attitude toward Light skin than Dark skin. However, our results also show that Albino skin has one of the lowest sentiment scores. The reasons for this difference may be twofold: First, as mentioned earlier, sentiment analysis is influenced by tweet volume, which may lead to biased results. Second, the previous study focused on price analysis but did not analyze the relevant social media sentiment or consider the impact of social sentiment involvement on CryptoPunks pricing. Therefore, there may be differences between the results of sentiment analysis and price analysis, and we will discuss these findings in more detail later in the analysis of CryptoPunks' pricing mechanics.

In conclusion, while these analyses cannot definitively conclude that discrimination occurred, these findings underscore the importance of recognizing and addressing these imbalances.

\subsection{The Influence of CryptoPunk Twitter Sentiment on CryptoPunk Valuation}
\label{sec:rq2}

After adding the sentiment score as an independent variable to the hedonic Regression model, we first test the correlation among all the independent variables and control variables. The correlation test results show that all the variables are weakly correlated, which allows us to use these variables in the hedonic regression model. Next, we conduct four regressions. In the first model, we only include skin tone and gender dummies as independent variables and include the rarity of CryptoPunks as a control variable. The second model adds three indicators of general market demand: daily growth in active wallets, growth in CryptoPunk sales volume in USD, and Ethereum gas price. The third model also adds ETH/USD exchange rate growth as a control variable. In the fourth model, we add the sentiment score as an independent variable. The base case is a Female CryptoPunk with an Albino skin tone.

According to the regression results (see \ref{tab:table1} in \ref{appendix:raw}), we find that in Model 2 and Model 3, after we added more control variables, the coefficients values of gender and skin tone only change slightly, and none of these coefficients change sign. The significance of each variable does not change, and all variables are statistically significant. After adding the sentiment score variable to the regression model, we find that the sentiment score is positively correlated with the price of CryptoPunks. This positive correlation is statistically significant. In addition, after adding the sentiment score, the light skin tone is also influenced, which becomes more statistically significant for price prediction than before. 

\begin{figure}[ht]
    \centering
    \includegraphics[width=0.4\textwidth]{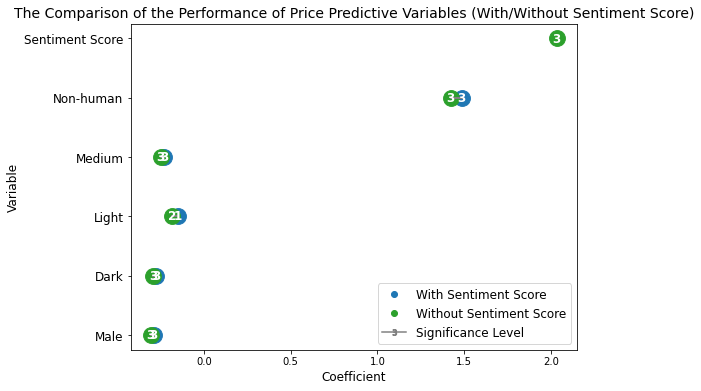}
    \caption{The comparison of the performance of price predictors (with and without sentiment score).}
    \label{fig:hrsentiment}
\end{figure}
To better illustrate the change in regression results before and after adding the sentiment score in the model, we put the coefficients of independent variables in \ref{fig:hrsentiment}. In the lollipop graph, each lollipop represents one independent variable, and the corresponding significance levels, which range from 0 to 3, are shown in the center of the lollipop. The larger the number, the more significant the variable, while 0 indicates no statistical significance. The green circles represent the model without the sentiment score, and the blue circles represent the regression with the sentiment score. From this graph, we could find the changing pattern more clearly.
 
Sentiments can significantly impact cryptocurrency prices. Wolk has used social media sentiment data to predict the price of cryptocurrency with least-square linear regression (LSLR) and Bayesian ridge regression models, and he concludes that people's attitudes toward cryptocurrency are very important in the price prediction of cryptocurrency \cite{wolk2020advanced}. Similarly, Tong, Goodell, and Shen quantify the impact of search engine attention (Google Trends) and social media attention (Twitter) on cryptocurrency returns through transfer entropy methodologies, and they conclude that social media sentiment is an important indicator of cryptocurrency price movements \cite{tong2022assessing}. We now verify that this situation holds for the CryptoPunk market; that is, sentiment is also an essential factor that influences the price of the CryptoPunk.

\subsection{The Structural Changes in CryptoPunk Valuation After 2021}
\label{sec:rq3}

\begin{figure}[ht]
    \centering
    \includegraphics[width=0.4\textwidth]{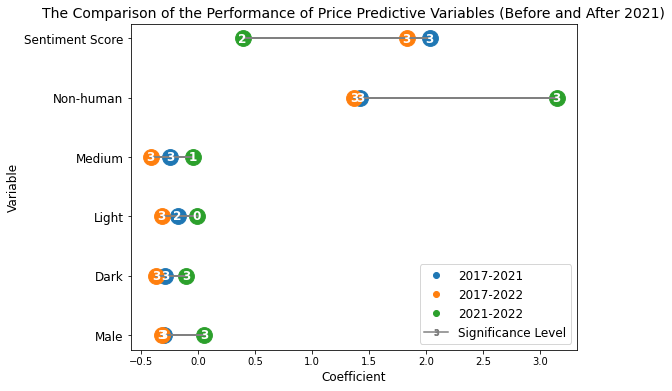}
    \caption{The comparison of the performance of price predictors (before and after 2021).}
    \label{fig:ba2021}
\end{figure}

As illustrated in \ref{sec:hedonic}, we expand our data from 2017-2021 to 2017-2022 (see \ref{tab:table2} in \ref{appendix:raw}). The comparison of the results of the regressions is shown in \ref{fig:ba2021}. We first conduct regression analysis on the data for 2021-2022. We compare these results with those from 2017 to 2021 and find significant changes in the impact of skin tone and gender on price predictions after 2021. For gender, we find a reversal of the previous trend, with males now predicted to have higher prices than females. For skin tones, the statistical significance of light skin is completely lost, and the statistical significance of medium skin on price prediction is greatly reduced. The nonhuman skin tone and sentiment scores still contribute positively to pricing. ~\footnote{Additionally, the coefficients of sentiment scores and Non-human CryptoPunks on prices show significant changes. We also conduct regression analysis on the overall data from 2017 to 2022, and the results are closer to those from before 2021. We attribute this to the amount of data, with the data from 2017 to 2021 accounting for around 80\% of the total data.}

In summary, our analysis suggests that there has been a structural change in the CryptoPunks valuation mechanics since 2021. Although our study has not focused on the reasons behind the fluctuation and structural changes, we may borrow the insights from a previous study that investigated the reasons for the rise and fall of a blockchain game---CryptoKitties \cite{jiang2021cryptokitties}. Jiang and Liu \cite{jiang2021cryptokitties} suggest that the fall of the game was due to (1) the oversupply of kitties, (2) the decrease in player income, (3) a widening gap between rich and poor players, and (4) the limitations of blockchain systems. Future research could investigate how the four aspects influence the valuation of CryptoPunks in particular and NFT in general.

\section{Conclusion}
\label{sec:conclusion}
Since reaching the peak of NFT transaction activities in late 2021, the NFT market has leveled off. However, studies on predictors of NFT prices after late 2021 are scarce, especially those involving ethical concerns, e.g., gender and skin tones. Examining the predictability of factors such as social media sentiment and ethics can help better understand the mechanics of NFT valuation and assess the ethical issues currently in the NFT marketplace. Therefore, in this study, we take the largest NFT collection, CryptoPunks, as an example, aiming to answer the questions related to the status quo of tweet sentiments, skin tones, and genders of CryptoPunks, as well as their predictability of CryptoPunk prices. Our research contributes to interdisciplinary study at the intersection of AI, ethics, and society, focusing on the ecosystem of decentralized AI or blockchain.\cite{zhang2023design} First, we examine the characteristics of the social and ethical factors of AI-generated AI--CryptoPunks. Second, we evaluated how sentiments in social media, together with gender and skin tone, contribute to NFT valuations by an empirical analysis of social media, blockchain, and crypto exchange data. Finally, we examine the structural changes in the mechanics of NFT valuation after transaction activities skyrocketed in 2021. Future research can further explore how NFT valuation relates to other factors in the crypto space such as the valuation of other crypto assets~\cite{liu2022deciphering, liu2022cryptocurrency,zhang2021optimal,zhang2022data}, the blockchain transaction fee mechanism~\cite{liu2022empirical, zhang2023understand}, and the decentralization level of blockchain transactions~\cite{zhang2022sok,ao_are_2022,zhang2022blockchain} or how does social media affect other aspects of blockchain economy including decentralization, security, and efficiency~\cite{zhuravskaya2020political, zhang2023design}.

\begin{acks}
    X. Tong and L. Zhang acknowledge the Data Science Research Center at Duke Kunshan University for funding their project entitled “A Data-Centered Approach to Investigate the Ethics Issues on Non-Fungible Tokens (NFT): Trust, Privacy, Transparency, and Fairness” under the Data + X Program. Luyao Zhang is supported by National Science Foundation China on the project entitled “Trust Mechanism Design on Blockchain: An Interdisciplinary Approach of Game Theory, Reinforcement Learning, and Human-AI Interactions (Grant No. 12201266).” We have benefited from the presentation and discussion at ChainScience Conference 2023, Boston, United States. (EasyChair paper ID: 16)
\end{acks}

\bibliographystyle{ACM-Reference-Format}
\bibliography{sample-base}

\appendix

\raggedbottom

\section{Hedonic Regression Statistics}\label{appendix:raw}

\begin{table}[H]
\centering
\resizebox{0.5\textwidth}{!}{
\begin{threeparttable}
\caption{Hedonic Regression Results from 2017 to 2021}
\label{tab:table1}
\begin{tabular}{l*{4}{c}}
\hline
\multicolumn{1}{l}{} & \multicolumn{4}{c}{\textbf{Dependent Variable: log(USD Price)}} \\
\hline
 & \textbf{(1)}  & \textbf{(2)} & \textbf{(3)} & \textbf{(4)} \\
\hline
\textbf{(Intercept)} & 8.5250$^{***}$ & 5.9371$^{***}$  & 5.9363$^{***}$  & 5.7096$^{***}$  \\
 & (0.1129) & (0.0912) & (0.0912) & (0.0914) \\
\textbf{Skin Tone: Dark} & -0.3917$^{***}$  & -0.2707$^{***}$  & -0.2735$^{***}$  & -0.2934$^{***}$  \\
& (0.1177) & (0.0880) & (0.0880) & (0.0871) \\
\textbf{Skin Tone: Light} & -0.2201* & -0.1467* & -0.1484* & -0.1827** \\
& (0.1184) & (0.0885) & (0.0885) & (0.0875) \\
\textbf{Skin Tone: Medium} & -0.4025$^{***}$  & -0.2288$^{***}$  & -0.2302$^{***}$  & -0.2497$^{***}$  \\
& (0.1177) & (0.0880) & (0.0880) & (0.0870) \\
\textbf{Skin Tone: Nonhuman} & 0.8535** & 1.5151$^{***}$  & 1.4861$^{***}$  & 1.4223$^{***}$  \\
& (0.3993) & (0.2984) & (0.2991) & (0.2958) \\
\textbf{Rarity} & 0.0000** & 0.0000$^{***}$  & 0.0000$^{***}$  & 0.0000$^{***}$ \\
& (0.0000) & (0.0000) & (0.0000) & (0.0000) \\
\textbf{Male} & -0.3871$^{***}$  & -0.2923$^{***}$  & -0.2906$^{***}$  & -0.3059$^{***}$  \\
& (0.0659) & (0.0493) & (0.0493) & (0.0488) \\
\textbf{Active Market Wallet (\% change, daily)} & & -0.0456$^{***}$  & -0.0506$^{***}$  & -0.0541$^{***}$ \\
& & (0.0034) & (0.0049) & (0.0049) \\
\textbf{CryptoPunk Sales Volume (USD, \% change, daily)} & & -0.0017$^{***}$  & 0.0030 & 0.0043 \\
& & (0.0003) & (0.0032) & (0.0032) \\
\textbf{Ethereum Gas Price (Gwei, average, daily)} & & 0.0000$^{***}$  & 0.0000$^{***}$  & 0.0000$^{***}$ \\
& & (0.0000) & (0.0000) & (0.0000) \\
\textbf{ETH/USD (\% change, daily)} & & & -0.0097 & -0.0125* \\
& & & (0.0067) & (0.0067) \\
\textbf{Sentiment Score} & & & & 2.0334$^{***}$  \\
& & & & (0.1348) \\
\hline
\textbf{R$^2$} & 0.0064 & 0.4457 & 0.4458 & 0.4581 \\
\textbf{Adjusted R$^2$} & 0.0058 & 0.4452 & 0.4452 & 0.4575 \\
\hline
\end{tabular}
\begin{tablenotes}[flushleft]
\small
\item Standard errors are in parentheses.

*$p<0.1$, **$p<0.05$, ***$p<0.01$.
\end{tablenotes}
\end{threeparttable}
}
\end{table}

\begin{table}[H]
\centering
\resizebox{0.5\textwidth}{!}{
\begin{threeparttable}
\caption{Hedonic Regression Results from 2021 to 2022}
\label{tab:table2}
\begin{tabular}{l*{4}{c}}
\hline
\multicolumn{1}{l}{} & \multicolumn{4}{c}{\textbf{Dependent Variable: log(USD Price)}} \\
\hline
 & \textbf{(1)}  & \textbf{(2)} & \textbf{(3)} & \textbf{(4)} \\
\hline
\textbf{(Intercept)} & 12.255$^{***}$  & 12.209$^{***}$  & 12.204$^{***}$  & 12.132$^{***}$  \\
 & (0.0301) & (0.0336) & (0.0331) & (0.0463) \\
\textbf{Skin Tone: Dark} & -0.1166$^{***}$  & -0.1201$^{***}$  & -0.1086$^{***}$  & -0.1063$^{***}$  \\
& (0.0311) & (0.0304) & (0.0299) & (0.0299) \\
\textbf{Skin Tone: Light} & -0.0138 & -0.0204 & -0.0141 & -0.0119 \\
& (0.0316) & (0.0310) & (0.0305) & (0.0305) \\
\textbf{Skin Tone: Medium} & -0.0571* & -0.0649** & -0.0540* & -0.0506* \\
& (0.0315) & (0.0308) & (0.0304) & (0.0304) \\
\textbf{Skin Tone: Nonhuman} & 3.2254$^{***}$  & 3.1366$^{***}$  & 3.1449$^{***}$  & 3.1484$^{***}$  \\
& (0.1499) & (0.1484) & (0.1461) & (0.1459) \\
\textbf{Rarity} & 0.0000** & 0.0000** & 0.0000$^{***}$  & 0.0000**\\
& (0.0000) & (0.0000) & (0.0000) & (0.0000) \\
\textbf{Male} & 0.0698$^{***}$  & 0.0605$^{***}$  & 0.0550$^{***}$  & 0.0532$^{***}$  \\
& (0.0184) & (0.0180) & (0.0178) & (0.0178) \\
\textbf{Active Market Wallet (\% change, daily)} & & -0.0257$^{***}$  & -0.0254$^{***}$  & -0.0243$^{***}$ \\
& & (0.0050) & (0.0050) & (0.0050) \\
\textbf{CryptoPunk Sales Volume (USD, \% change, daily)} & & -0.0112** & 0.0092** & 0.0089* \\
& & (0.0047) & (0.0046) & (0.0046) \\
\textbf{Ethereum Gas Price (Gwei, average, daily)} & & 0.0000$^{***}$  & 0.0000$^{***}$  & 0.0000$^{***}$ \\
& & (0.0000) & (0.0000) & (0.0000) \\
\textbf{ETH/USD (\% change, daily)} & & & 1.4900$^{***}$  & 1.4885$^{***}$  \\
& & & (0.2154) & (0.2151) \\
\textbf{Sentiment Score} & & & & 0.3885** \\
& & & & (0.1748) \\
\hline
\textbf{R$^2$} & 0.2731 & 0.3073 & 0.3293 & 0.3316 \\
\textbf{Adjusted R$^2$} & 0.2701 & 0.3030 & 0.3247 & 0.3265 \\
\hline
\end{tabular}
\begin{tablenotes}[flushleft]
\small
\item Standard errors are in parentheses.

*$p<0.1$, **$p<0.05$, ***$p<0.01$.
\end{tablenotes}
\end{threeparttable}
}
\end{table}

\end{document}